\documentclass[aps,prb,twocolumn,groupedaddress]{revtex4-1}

\usepackage{graphicx}
\usepackage{mathtools,amsthm,amssymb,braket}
\usepackage{bm}
\usepackage[utf8]{inputenc}
\usepackage[german,  danish, english]{babel}
\usepackage{siunitx}

\usepackage[pdfauthor={V. O. Shkolnikov, and Guido Burkard},
pdftitle={Theory of generalized Lambda systems},
pdfsubject={},colorlinks=true,linkcolor=blue,citecolor=magenta]{hyperref}

\usepackage{floatflt}

\renewcommand{\d}{{\mathrm{d}}}

\DeclareSIUnit\gauss{G}

\begin{document}

\title{Magnetic Resonance in Defect Spins mediated by Spin Waves}
\author{Clara M\"uhlherr, V. O. Shkolnikov, and Guido Burkard}
\affiliation{Department of Physics, University of Konstanz, D-78464 Konstanz, Germany}


\begin{abstract}
In search of two level quantum systems that implement a qubit, the nitrogen-vacancy (NV) center in diamond has been intensively studied for years. Despite favorable properties such as remarkable defect spin coherence times, the addressability of NV centers raises some technical issues. The coupling of a single NV center to an external driving field is limited to short distances, since an efficient coupling requires the NV to be separated by only a few microns away from the source. As a way to overcome this problem, an enhancement of coherent coupling between NV centers and a microwave field has recently been experimentally demonstrated using spin waves propagating in an adjacent yttrium iron garnet (YIG) film\cite{Andrich}. 
In this paper we analyze the optically detected magnetic resonance spectra that arise when an NV center is placed on top of a YIG film for a geometry similar to the one in the experiment. We analytically calculate the oscillating magnetic field of the spin wave on top of the YIG surface to determine the coupling of spin waves to the NV center. We compare this coupling to the case when the spin waves are absent and the NV center is driven only with the antenna field and show that the calculated coupling enhancement is dramatic and agrees well with the one obtained in the recent experiment.
\end{abstract}

\pacs{}

\maketitle

\section{\label{Sec1}Introduction}
The negatively charged nitrogen vacancy (NV) center in diamond is an optically active point defect with a ground state spin triplet, lying deep in the bandgap of diamond \cite{1367-2630-13-2-025025,DOHERTY20131}. Its bright optical transition and the existence of intersystem crossing provides a good mechanism for initialization and read out of the spin state of the center\cite{PhysRevLett.92.076401}. The ground state of the NV center is sensitive to magnetic and electric fields, as well as to strain \cite{doi:10.1021/acs.nanolett.6b04544, PhysRevB.85.205203, PhysRevB.98.075201}, and thus can be used as a nanosensor to detect them\cite{Balasubramanian2008, Maze2008, Barfuss2015}. This makes the NV center extremely interesting for metrology. Apart from that, the long spin coherence time of the defect makes it interesting for quantum information purposes. Its state can efficiently be manipulated with oscillating magnetic fields, that cause transitions between the levels of the spin triplet. In most of the experiments this oscillating field was generated by an antenna placed in the vicinity of the center, which raises the issue of addressability for many NV centers when the antenna can no longer be placed close to each of them. Recently an experiment was reported in which the NV center was placed on top of a ferromagnetic material (YIG) that can host propagating spin waves. Using an antenna as a source of the spin waves in this material, one can couple distinct NV centers to it just like dialog partners are connected by a signal line. 
In this work we theoretically treat the coupling of the spin waves to the NV centers for a special geometry of the device described below. We provide an analytical expression for the spin wave field and determine the coupling enhancement that it produces with respect to the field of the antenna only, when spin waves are absent.

\begin{figure*}
\includegraphics[width=1\textwidth]{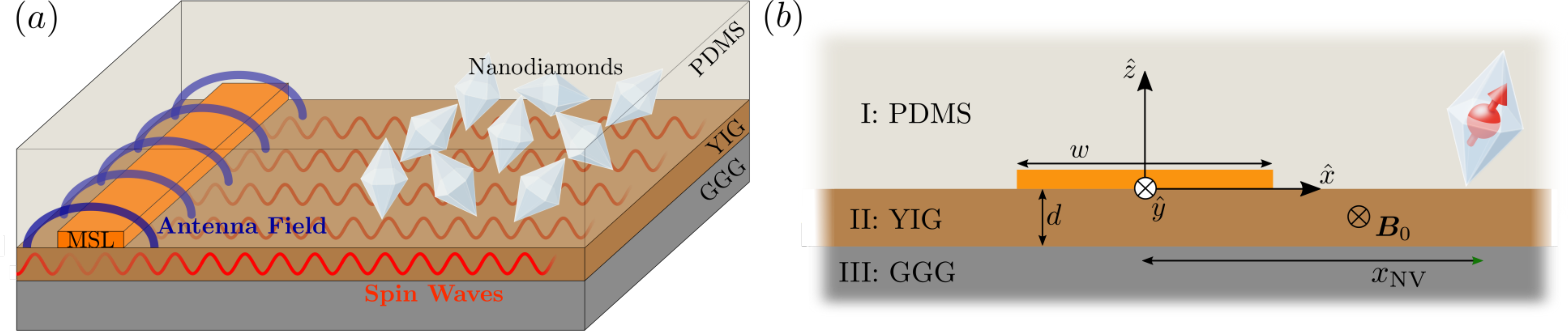}   
\caption{$(a)$ Electron spin resonance in NV spins driven by spin waves demonstrated by Andrich \textit{et al.} \cite{Andrich}. An array of nanodiamonds is patterned inside a PDMS film layered on top of a YIG thin film. From a distant MSL grown on the YIG the antenna field propagates through the PDMS and couples to the NV centers. Inside the YIG the microwave excitation by the MSL leads to spin waves propagating in the plane. Due to the given dimensions, the theoretical treatment can be reduced to a two-dimensional coordinate system as shown in $(b)$, which is a cross section of the setup in $(a)$. The layered structure lies in the $xy$-plane and the center of the MSL of width $w$ marks the origin of the coordinate system and is oriented along the $y$-direction.
\label{fig:Setup} }
\end{figure*}

\section{Setup \label{Sec2}}
We calculate the effect of spin wave excitation inside ferromagnetic thin films as used for example in the experiment presented by Andrich \textit{et al.}\cite{Andrich}. The vicinity of a spin wave hosting material to a quantum system driven by external microwaves provides an additional component of the driving field. In that way, spin wave excitation inside a ferromagnetic material affects the coupling between the driving field and the given quantum system.
In the experimental realization by Andrich et al.\cite{Andrich} the quantum system consisted of defect spins inside a collection of nanodiamonds patterned on the surface of the magnetic material as depicted in Fig. \ref{fig:Setup}$(a)$.
A single nanodiamond, which is selected with the laser focus, hosts an ensemble of $\sim500$ NV centers with an isotropic electron g-factor of $g\approx 2$  \cite{DOHERTY20131}. The nanodiamond is embedded in a polydimethylsiloxane (PDMS) film, which is on the top of a layered structure of YIG and gadolinium gallium garnet (GGG). Since YIG is a ferromagnetic material with ultra-low spin wave damping, it is perfect for the usage as the spin wave medium. An ac current flowing through a microstrip line (MSL) grown on the surface of the YIG generates a microwave driving field, penetrating through the different materials and denoted as the antenna field in Fig. \ref{fig:Setup}$(a)$. Due to the vicinity of the NV centers to the YIG surface, the defect spins are not only sensitive to the antenna field, but also to the stray field originating from the magnetic YIG film. 
For a theoretical description of the system we use the coordinate system and the defining parameters sketched in Fig. \ref{fig:Setup}$(b)$.
Choosing the YIG film to have a thickness $d$ and to lie in the $xy$-plane, the MSL orientation can be set as the $y$-axis and the MSL position marks the origin of the coordinate system. Further, the MSL has a width $w$ and the probed nanodiamond is located in the $xz$-plane at $x_\text{NV}$. Based on realistic dimensions of experimental systems, some assumptions concerning the boundary conditions of the system are reasonable. In Ref.~\onlinecite{Andrich}, the YIG film is only $\SI{3.08}{\micro\meter}$ thick, which is very thin compared to its dimensions of about $\SI{10}{\milli\meter}$ along the $x$- and $y$-axis\cite{Andrich}. Thus, the film can be assumed to be infinite in the $xy$-plane. The GGG substrate and the PDMS layer are a few hundreds of $\si{\micro\meter}$ thick, so that they are treated as infinite in the $z$-direction and the only boundary conditions to be fulfilled are those at the YIG interfaces, where both surrounding layers in regions I and III are approximated as non-magnetic. This assumption is justified due to their magnetic permeabilities being isotropic and close to 1\cite{GGGPerm}\cite{PDMSPerm}. The MSL has a width of $w=\SI{5}{\micro\meter}$, making its height of about $\SI{200}{\nano\meter}$ negligible.
Overall, due to the spatial expansion of the system along the $y$-direction and the invariance of the system under $y$-translation, we assume that all fields are independent of  $y$ and the problem is treated in two dimensions.
\section{\label{Sec3}Driven Spin Waves}
In order to calculate the spin wave spectrum of a ferromagnetic thin film and the resulting field amplitude, we start from Maxwell's equations (MEs). If there is an external magnetic $\bm{H}$-field, a magnetization field $\bm{M}$ is built up in the magnetic film. In general, the external $\bm{H}$-field can be decomposed into a static part $\bm{H_0}$ and a time-dependent component $\bm{h}(t)$
originating from the microwave antenna field. Consequently, the magnetization inside the material depends on the driving frequency $\omega$ and will also have a time-dependent component $\bm{m}(t)$. The relation between both time-varying components is given by the constitutive equation $\bm{m}=X\bm{h}$, where material properties and the geometry of the system  enter via the susceptibility tensor $X$. In case of a strong static bias field $\bm{H_0}$ along the $y$-direction, the magnetic film is tangentially magnetized and the static component of magnetization $\bm{M_0}$ saturates. Under these conditions, $X$ takes the form of the Polder susceptibility 
\begin{equation}\label{eq:Polder}
X=\begin{pmatrix}
\chi & 0 & -i\kappa\\
0&1&0\\
i\kappa & 0 & \chi\\
\end{pmatrix},
\end{equation}
with the frequency-dependent entries $\chi=\omega\omega_M/(\omega_0^2-\omega^2)$ and $\kappa=\omega_0\omega_M/(\omega_0^2-\omega^2)$\cite{Stancil}. The parameters $\omega_0=\gamma\mu_0H_0$ and $\omega_M=\gamma\mu_0M_S$ account for the characteristics of the material in an external field $H_0$.
Here, $\gamma$ and $M_S$ denote the gyromagnetic ratio and the saturation magnetization of the film and $\mu_0$ denotes the vacuum permeability.
Using the Polder tensor \eqref{eq:Polder} and assuming the electric permittivity of the materials to be 1 for simplicity, the single components of the four MEs in two dimensions form a system of eight coupled differential equations for the magnetic and electric field components $H_i(x,z)$ and $E_i(x,z)$ ($i=x,y,z$) in each region I-III in Fig. \ref{fig:Setup}$(b)$.
Since the MSL is located right at the PDMS-YIG interface and is assumed to be infinitesimally thin, the flowing current is non-zero only at the boundary between the two upper layers I and II, whereas inside the bulk regions there are no free currents $\bm{j}$ flowing, and thereby, no additional source terms. Referring to that, the fields inside the bulk regions I-III are obtained by solving the system of homogeneous MEs with $\bm{j}=0$ separately and the MSL current is included by matching the boundary conditions at $z=0$ and $z=-d$ afterwards.

Performing a one dimensional Fourier transform of the $x$-coordinate yields a system of ordinary differential equations in the $k_xz$-space, where only one equation actually has to be solved,
\begin{equation}\label{eq:BasicDiffEq}
\partial_z^2 H_x(k_x,z)+a^2 H_x(k_x,z)=0,
\end{equation}
where $a^2=((1+\chi)^2-\kappa^2)k_0^2/(1+\chi)-k_x^2$ and $k_0=\omega/c$. The other non-zero field components $H_z(k_x,z)$ and $E_y(k_x,z)$ can be expressed in terms of the solution $H_x(k_x,z)$ of \eqref{eq:BasicDiffEq} and its derivative $\partial_zH_x(k_x,z)$,
\begin{align}
\label{eq:Hy(Hx)}
H_z(k_x,z)&=\frac{i\kappa k_0^2 H_x(k_x,z)-i k_x\partial_z  H_x(k_x,z)}{k_x^2-(1+\chi)k_0^2},\\
\label{eq:Ey(H)}
E_y(k_x,z)&=\frac{\omega}{k_x}(i\kappa H_x(k_x,z)+(1+\chi)H_z(k_x,z).
\end{align}
Since the PDMS as well as the GGG layer are assumed to be infinite in positive and negative $z$-direction, there are no incoming waves in these regions, which could be caused by reflections at any surfaces. Thus, the ansatz
\begin{equation}\label{eq:AnsatzHx}
\begin{aligned}
H_x^\text{I}(k_x,z)=&~~C_1~e^{ia^\text{PDMS}z},\\
H_x^\text{II}(k_x,z)=&~~C_2~e^{ia^\text{YIG}z}+C_3~e^{-ia^\text{YIG}z},\\
H_x^\text{III}(k_x,z)=&~~C_4~e^{-ia^\text{PDMS}z},
\end{aligned}
\end{equation}
is chosen, where the superscripts I-III refer to the regions and $a^i$ corresponds to $a$ in \eqref{eq:BasicDiffEq} in the corresponding material $i$.
In order to obtain the actual amplitude of the magnetic field, the coefficients $C_1$-$C_4$ have to be derived, so there is need to include existing boundary conditions, which arise at the two interfaces. In the absence of any surface currents, the parallel component of the $\bm{H}$-field and the orthogonal component of the $\bm{B}$-field are continuous at an interface. 
But since this is only the case for the lower interface at $z=-d$, the upper boundary condition for the parallel $\bm{H}$-field component has to be considered more carefully. The field component $H_x$ is not continuous at $z=0$, where the step between both regions equals the current density at the boundary. Hence, the proper boundary condition at the I-II interface is
\begin{equation}\label{eq:BoundaryCondition}
H_x^\text{I}\left(k_x,0\right)-H_x^\text{II}\left(k_x,0\right)=j(k_x,y)
\end{equation}
in the $k_xz$-space. The current density function $j(x,z)$ describes the total current $I_0$ flowing through the infinitesimally thin MSL of width $w$, what can be expressed as $j(x,z)=I_0/w\vartheta(w/2-x)\vartheta(x+w/2)\delta(z)$ and the corresponding Fourier transform is $j(k_x,z)= j_0 \sin(k_xw/2)/k_x$ with $j_0=(2/\pi)^{1/2}I_0/w$. Combining this discontinuity of the parallel $\bm{H}$-field at the upper interface with the known continuity condition at the lower interface and the continuity of the orthogonal $\bm{B}$-field at both interfaces provides four boundary conditions in total. Inserting the ansatz \eqref{eq:AnsatzHx} finally leads to a system of linear equations determining the coefficients $C_1-C_4$. As long as the interacting quantum system is located above the magnetic thin film, we deal require the solution for positive $z$-values, so that only the field amplitude in region I is playing a role for any coupling processes. Hence, it is sufficient to concentrate only on the coefficient $C_1$.
The system of equations can be further simplified by introducing approximations relying on the realistic experimental values\cite{Andrich}. A typical saturation magnetization $M_S$ of the ferromagnetic material is of the order of $10^3~\si{\gauss}$ and is achieved in an external magnetic field $B_0$ of $\sim 10^2~\si{\gauss}$. Hence, the assumption $\omega_0^2\ll\omega^2$ is justified for microwave excitation frequencies $\omega$ in the $\si{\giga\hertz}$ range and susceptibility parameters $\chi$ and $\kappa$ of the order of $10^{-1}$ and $10^0$. Further, the wave vectors of microwaves of about $k_0\approx 10^1~\si{\per\meter}$ in vacuum are much smaller than the experimental wave vectors $k_x\approx10^5\si{\per\meter}$, i. e. $k_0^2\ll k_x^2$. Under these conditions, the parameter $a$ in \eqref{eq:BasicDiffEq} can be approximated independently from the material as $a^\text{PDMS}\approx a^\text{GGG}\approx a^\text{YIG}\approx ik_x$. Hence, the result for $C_1$ only depends on $k_x$,
\begin{widetext}
\begin{equation}\label{eq:C1(kx)}
C_1(k_x)\approx j_0~ \frac{\sin(k_xw/2)}{k_x} ~\frac{\omega_M(\omega_0+\omega_M+\omega)-e^{2k_xd}(\omega_0+\omega_M-\omega)(2\omega_0+\omega_M+2\omega)}{\omega_M^2+e^{2k_xd}(4\omega^2-(2\omega_0+\omega_M)^2)}.\vspace*{1cm}
\end{equation}
\end{widetext}
\begin{figure*}

\includegraphics[height=0.18\textheight]{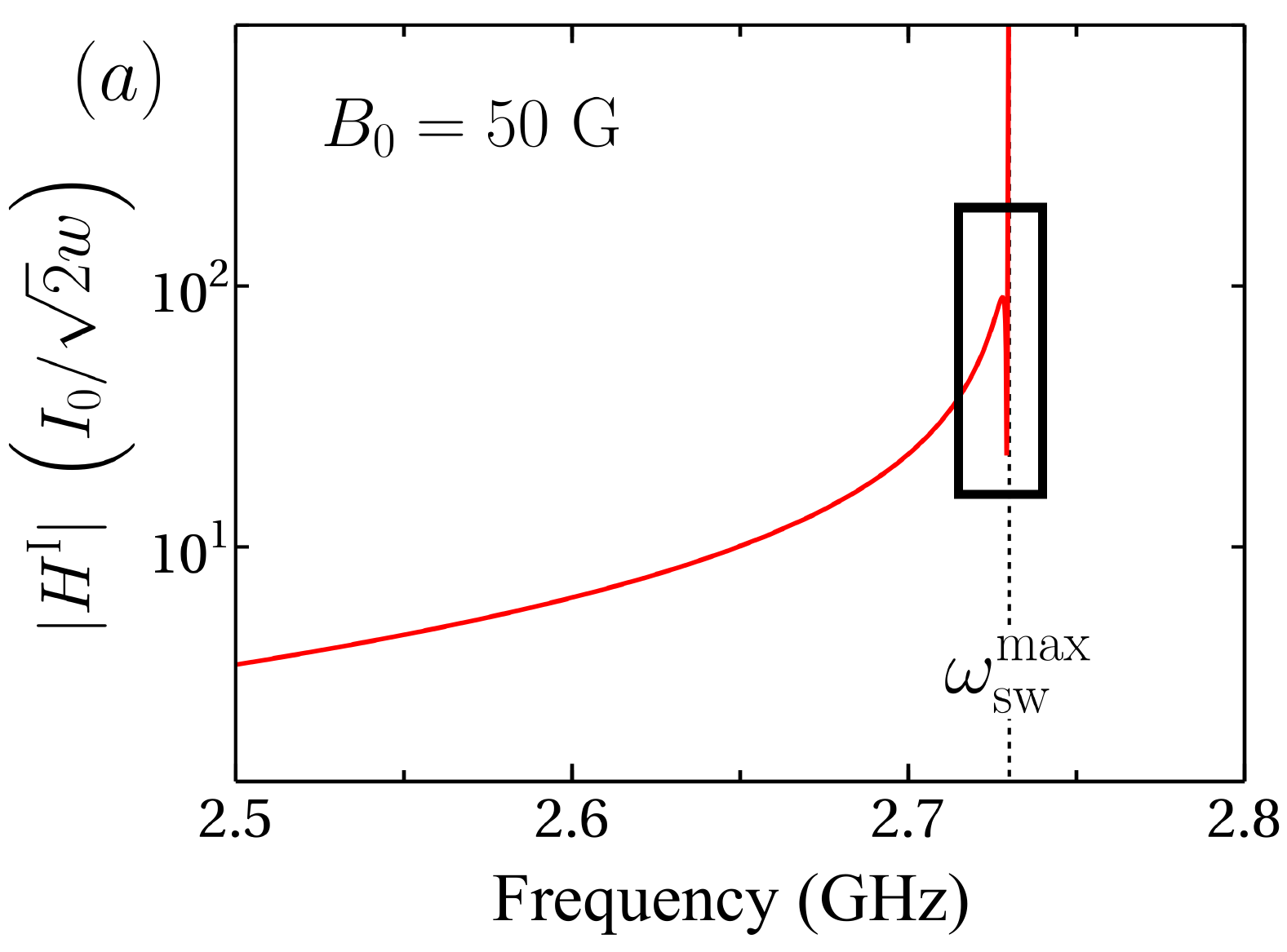} \includegraphics[height=0.18\textheight]{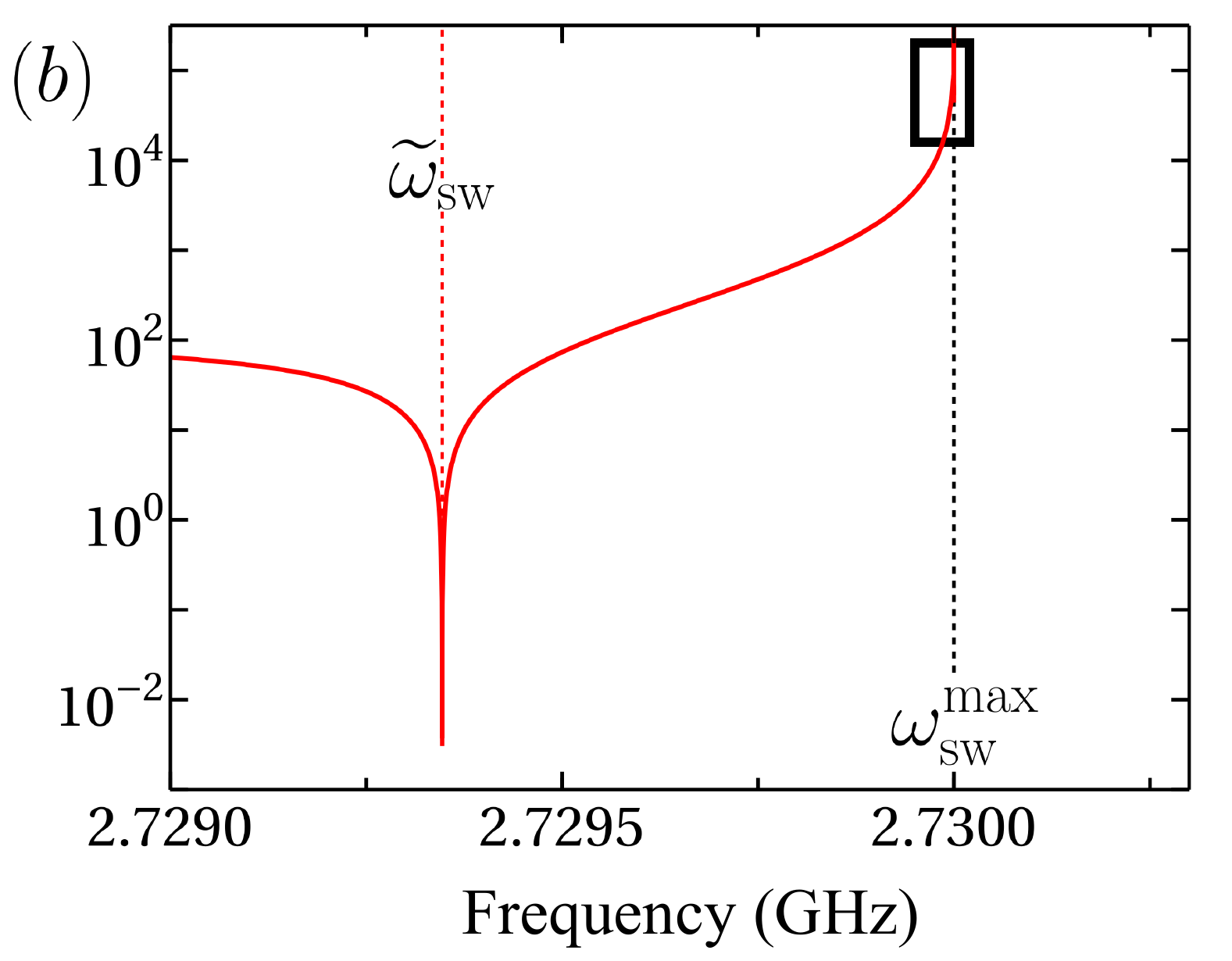}
\includegraphics[height=0.18\textheight]{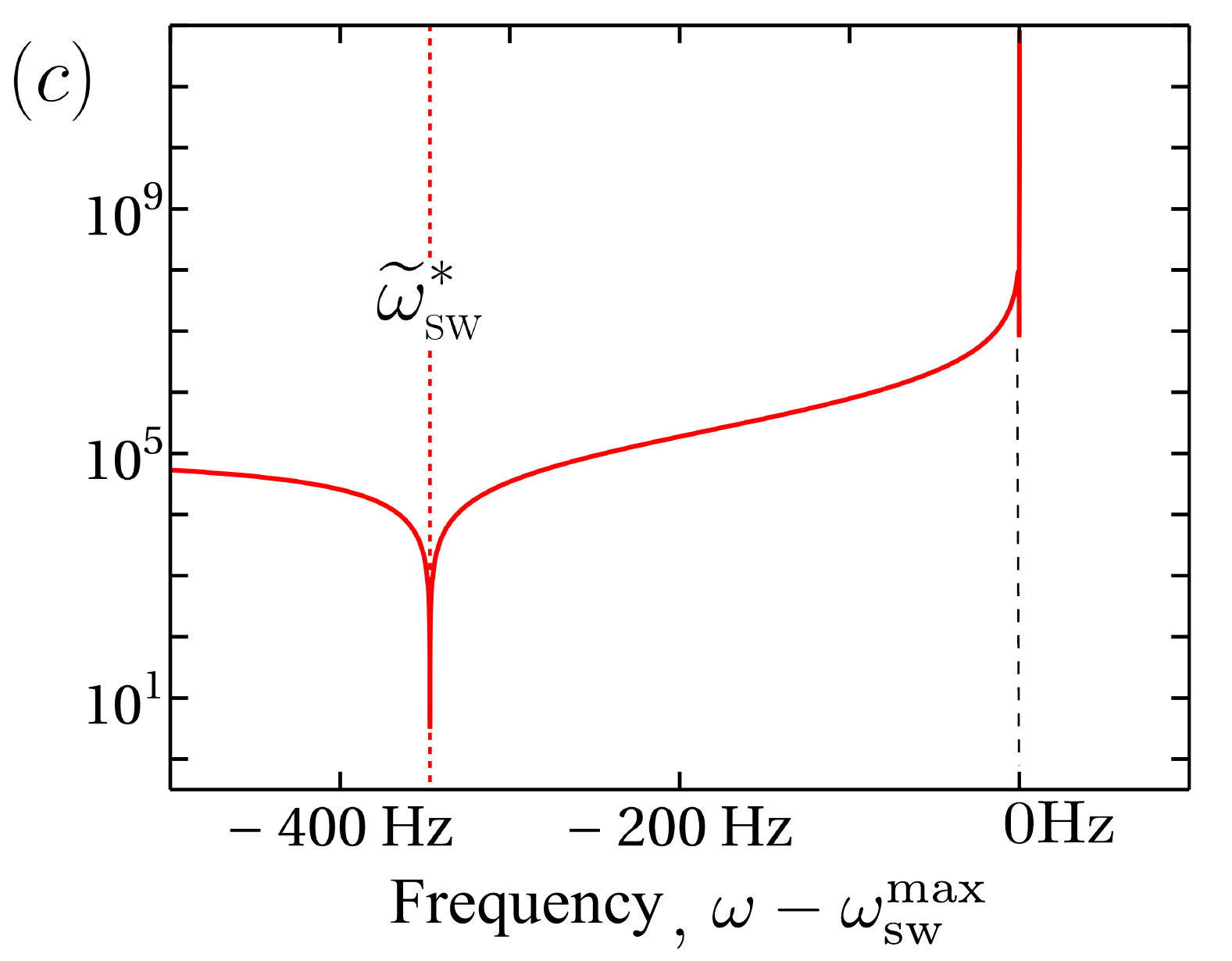}
   \caption{Magnetic field amplitude depending on the microwave frequency $\omega$ at fixed external field $B_0=\SI{50}{\gauss}$. The field amplitudes are normalized to the applied current per unit length $I_0/w$. $(a)$ The wide range plot shows the step behavior at the upper bound $\omega_\text{sw}^\text{max}$ of spin wave excitation. $(b)$ Magnification of the range around the step frequency indicated by the black rectangle in $(a)$, which resolves a sharp decrease in the amplitude at $\widetilde{\omega}_\text{sw}$. $(c)$ Further magnification shows a second amplitude dip at $\widetilde{\omega}^\ast_\text{sw}$, below $\omega_{\text{sw}}^{\text{max}}=\SI{2.73}{\giga\hertz}$. 
\label{fig:C1Plot}}
\end{figure*}
The full solution for the magnetic field $\bm{H}$ in the region I is given by 
\begin{equation}\label{eq:MagneticFieldkx}
\begin{pmatrix}
H_x^\text{I}\\[0.1cm]H_z^\text{I}
\end{pmatrix}
\approx\begin{pmatrix}
1\\i
\end{pmatrix}C_1(k_x)~e^{-k_x z}\equiv\begin{pmatrix}
1\\i
\end{pmatrix}H^\text{I}(k_x,z)/\sqrt{2},
\end{equation}
where the amplitude function $H^\text{I}(k_x,z)$ is introduced. From equation \eqref{eq:MagneticFieldkx} as a function of $k_x$ and $z$, the excited spin wave modes inside the ferromagnetic film, i. e. the spin wave resonance condition, can be derived by determining the zeroes of the denominator. 
This yields the dispersion relation 
\begin{equation}\label{eq:DispDESW}
\omega(k_x)=\frac{1}{2}\sqrt{(2\omega_0-\omega_M)^2-\omega_M^2e^{-2k_xd}},
\end{equation}
that matches exactly the so called Damon-Eshbach surface waves (DESW), which were calculated for ferromagnetic thin films in the absence of any surface currents\cite{DamonEshbach}. Hence, the solution of the modified system with non-zero current is peaked around the DESW modes, which is reasonable against the background of made approximations.
Assuming large wave numbers $k_x$, i. e. $k_x\gg k_0$, the current function $j(k_x,z) \sim \sin(k_x w/2)/k_x$ decreases in amplitude, so that the result for the resonance condition is identical to the DESW modes within the given approximation.

An important feature of the calculated spin wave modes is that for large $k_x$ the frequency in \eqref{eq:DispDESW} saturates to a value that depends on the external magnetic field $B_0$.
Thus, at a fixed magnetic field, spin wave excitations are limited to a fixed range of frequencies between the limiting cases $k_x=0$ and $k_x\rightarrow \infty$,
\begin{equation}\label{eq:FrequencyRange}
\sqrt{\omega_0(\omega_0-\omega_M)}~\leq~ \omega~\leq~ \omega_0-\omega_M/2.
\end{equation}
The lower bound at $k_x=0$ corresponds to the so called uniform precession mode, where all the spins inside the material precess in phase, so that there is an oscillating magnetization, but no spatial propagation. Even in a strong external magnetic field of $\SI{200}{\gauss}$ as used, for example, in the reported experiment\citep{Andrich}, the uniform precession mode of the considered YIG film oscillates at about $\SI{1.6}{\giga\hertz}$, which is far too small to stimulate magnetic dipole transitions in the spin triplet of the NV center. In contrast, the upper limit $\omega_\text{sw}^\text{max}=\omega_0-\omega_M/2$ is reached for the same magnetic field at $\sim\SI{3.1}{\giga\hertz}$, which lies, for example, in the range of NV resonances, as will be important later.

After deriving the driven spin wave modes, we are interested in the real space solution of the magnetic field, to finally model the interaction with an NV ensemble and to give a quantitative expectation of the coupling strength. Therefore, the magnetic field $\bm{H}^\text{I}$ in \eqref{eq:MagneticFieldkx} has to be Fourier transformed back into real space using the explicit form of the coefficient $C_1(k_x)$ in \eqref{eq:C1(kx)}. The field can be written as
\begin{equation}\label{eq:MagneticFieldX}
\bm{H}^\text{I}(x,z)=\begin{pmatrix}
1\\i
\end{pmatrix}H^\text{I}(x,z)/\sqrt{2}
\end{equation}
with the Fourier transform of the amplitude function 
\begin{equation}\label{eq:C1(x)1}
H^\text{I}(x,z)=
\frac{1}{\sqrt{2\pi}}\int_{-\infty}^\infty H^\text{I}(k_x,z)~e^{ik_x x}~\d k_x.
\end{equation} 
In order to calculate this integral, the denominator in \eqref{eq:C1(kx)} is expanded to first order in $k_x$ around its zero $k'_x$, which corresponds to the resonant wave number defined by \eqref{eq:DispDESW},
\begin{equation}
k'_x=\frac{1}{2d}~\ln\left(-\frac{\omega_M^2}{4\omega^2-(2\omega_0+\omega_M)^2}\right).
\end{equation}

The remaining integral can be analytically evaluated by assuming that before the current in the MSL was switched on there were no spin waves. This imposes the rule how to go around the pole in the integral above. Applying the Sokhotski$\text{-}$Plemelj theorem in case of a real line integral and restricting ourselves to the far field regime, where the condition $x\gg(k_x-k'_x)^{-1}$ holds, the amplitude function $H^\text{I}(x,z)$ in real space is
\begin{widetext}
\begin{equation}\label{eq:C1(x)Result}
H^\text{I}(x,z)\approx i\sqrt{2}~\frac{I_0}{w}~\frac{\sin(k'_xw/2)}{k'_xd}\left[\frac{\omega_M(\omega_0+\omega_M+\omega)}{4\omega^2-(2\omega_0+\omega_M)^2}~e^{-2k'_xd}+\frac{\omega_0+\omega_M-\omega}{2\omega_0+\omega_M-2\omega}\right]e^{-k'_xz}~e^{ik'_xx},
\end{equation}
\end{widetext}
which is one of our main results. Based on this complex field amplitude, the magnetic field above the ferromagnetic film is known explicitly. Note, however, that the calculated solution only holds for frequencies within the range \eqref{eq:FrequencyRange}. The assumptions, which were made in order to solve the integral above, are not valid outside this frequency range, where the existence of the solution \eqref{eq:C1(x)Result} at frequencies above $\omega_\text{sw}^\text{max}$ is not given. Hence, the given solution \eqref{eq:C1(x)Result} is only valid for $\sqrt{\omega_0(\omega_0-\omega_M)}\leq \omega\leq \omega_0-\omega_M/2$ and is set to 0 otherwise, which corresponds to the absence of spin waves.
\begin{figure*}
\includegraphics[height=0.2\textheight]{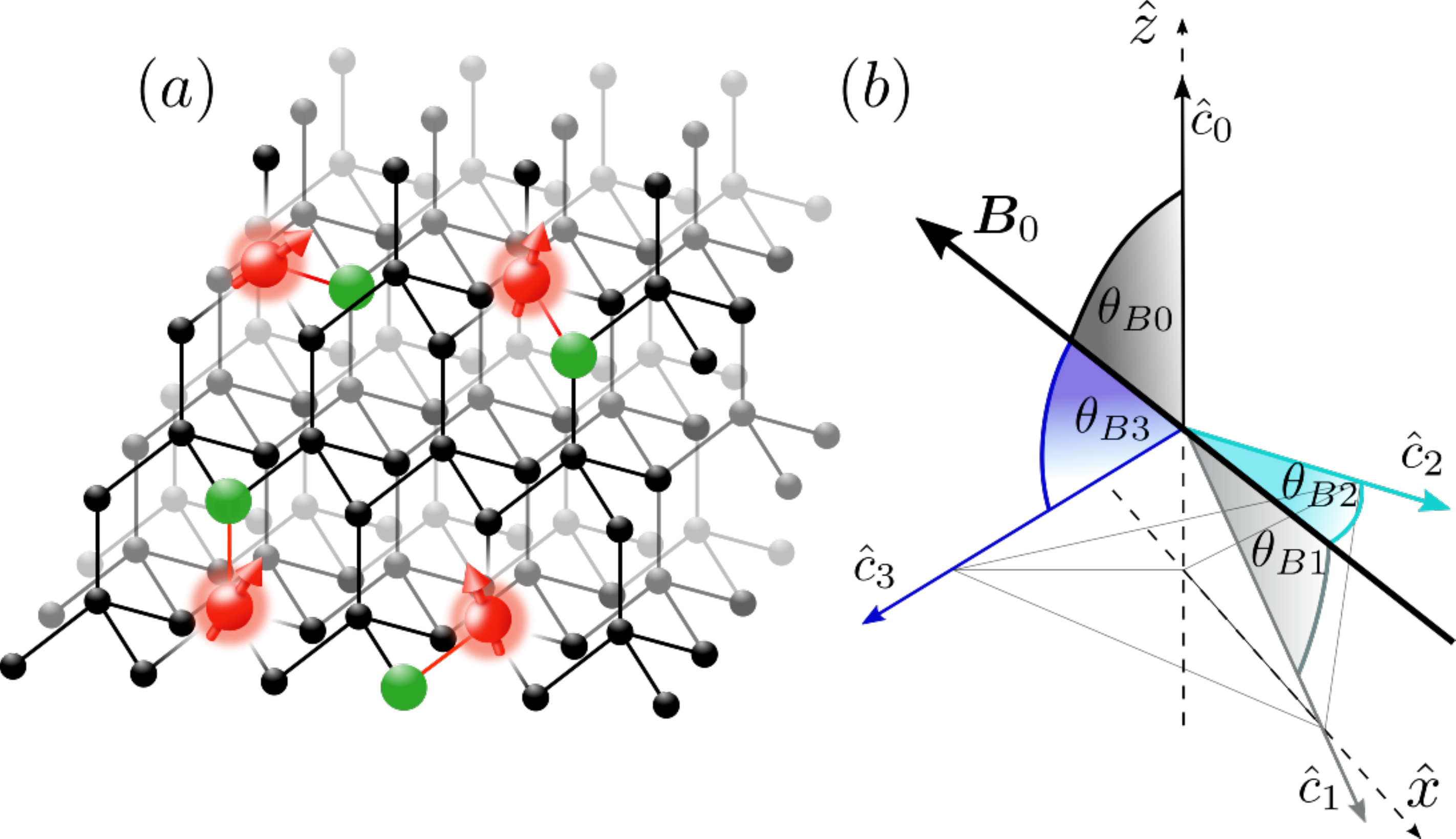}\hfill \includegraphics[height=0.2\textheight]{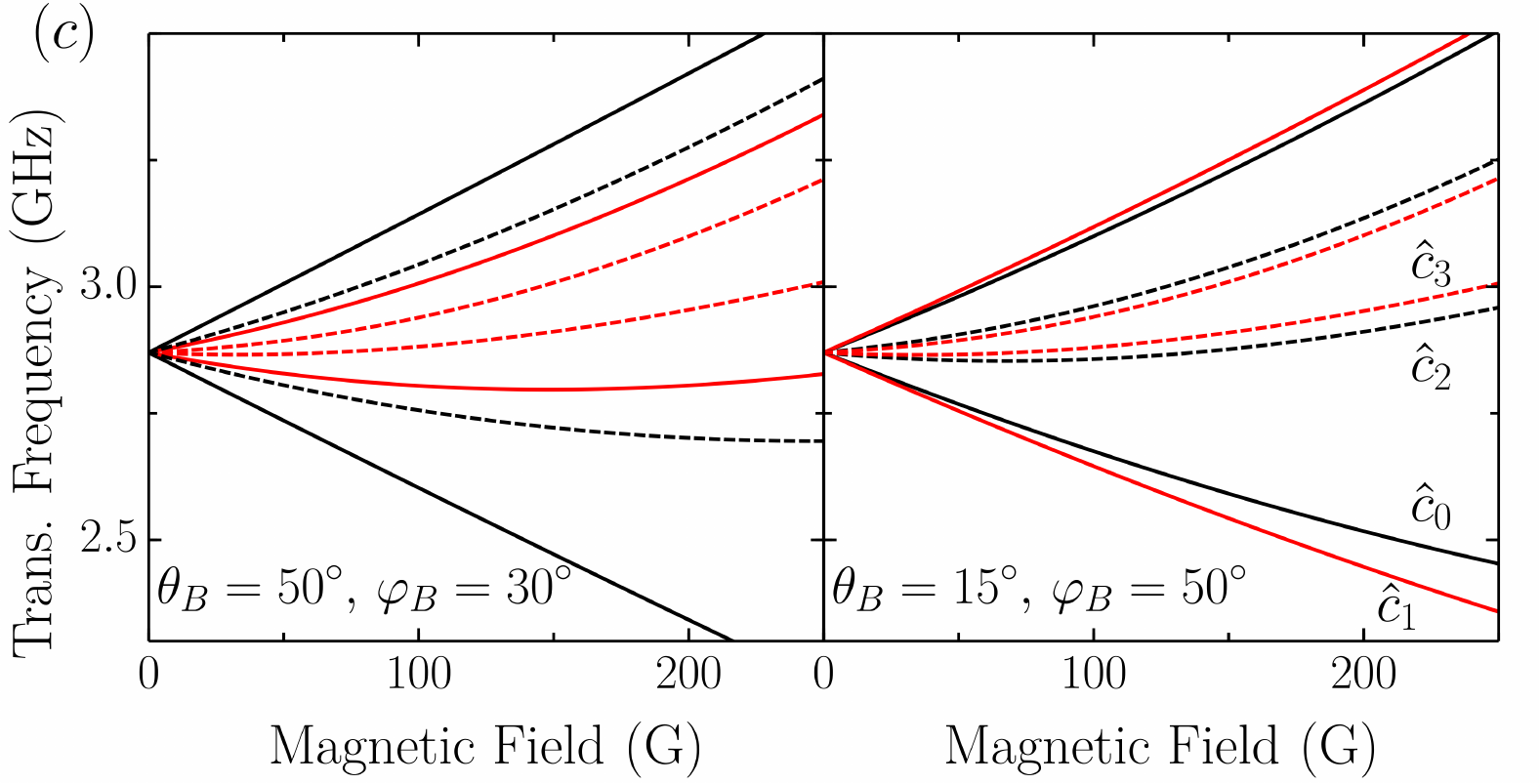}
   \caption{$(a)$ Possible orientation of NV centers inside the diamond structure. Depending on the alignment of the connection line between a substitutional nitrogen atom (green) and the neighboring vacancy (red), the NV-axis (red) can be oriented in the depicted four directions. $(b)$ The crystal structure of diamond leads to four different angles $\theta_{Bi}$ between the magnetic field and the possible NV-axes $\hat{c}_i$. $(c)$ Resonance frequencies of the magnetic dipole transitions within the ground state triplet of an ensemble of NV centers embedded in single crystal diamond. Each of the four possible NV center orientations $\hat{c}_i$ contributes two resonances at $\omega^\pm$ indicated by the different line styles. The resonances depend on the alignment of the magnetic field. For the left plot in $(c)$ the magnetic field is $\theta_B=\SI{50}{\degree}$ at an angle $\varphi_B=\SI{30}{\degree}$ with respect to one of the crystal axes and on the right the chosen parameters are  $\theta_B=\SI{15}{\degree}, \varphi_B=\SI{50}{\degree}$.
   }
\label{fig:NVCrystalAngle}
\end{figure*}
Within this definition range, the absolute value of the field amplitude in \eqref{eq:C1(x)Result} at fixed magnetic field depends on the excitation frequency as plotted in Fig. \ref{fig:C1Plot}. The wide range plot in Fig. \ref{fig:C1Plot}$(a)$ shows the expected behavior. As discussed before, an upper bound at $\omega_\text{sw}^\text{max}$ cuts off the spin wave driving regime. Above this limit, the amplitude rapidly drops to a level at least six orders of magnitude smaller. Below this limit, the  amplitude increases until the resonance frequency $\omega_\text{sw}$ for DESW modes in \eqref{eq:DispDESW} is reached, which becomes apparent as a high peak just below the cut-off frequency.
Although the increase of $|H^\text{I}|$ from low frequencies towards $\omega_\text{sw}^\text{max}$ in Fig. \ref{fig:C1Plot}$(a)$ seems to be smooth at the first sight, the magnifications in $(b)$ and $(c)$ highlight a substructure of amplitude dips very close to the maximum. The occurring amplitude dips correspond to the minima of the amplitude function in \eqref{eq:C1(x)Result} induced by the zeros of the factor $\sin(k_x'w/2)/k'_x$. The position of the amplitude dip at $\widetilde{\omega}^\ast_\text{sw}$ in $(c)$ differs by only a few hundreds of Hz from the resonance maximum and, therefore, might not be resolved in usual resonance experiments. In contrast, the second dip in $(b)$ is more incisive and occurs at a frequency a few MHz from the maximum. This second decrease in the field amplitude should be experimentally resolvable as discussed later in section \ref{Sec5}.

\section{\label{Sec4}Magnetically driven NV spins}
The spin wave propagation through the magnetic thin film can increase the interaction between the external field and a quantum system. As a consequence, the driving of magnetic dipole transitions of NV centers in diamond can be performed more efficiently. In order to model the coupling strength and to simulate the resulting transition spectrum, we have to distinguish whether a single NV center or an NV ensemble in single crystal diamond is probed.
\subsection{Single NV}
The ground state of the NV center is a spin triplet, which is split into the $m_S=0$ and $m_S=\pm1$ sublevels. This ground state energy splitting $D$ of $\SI{2.87}{\giga\hertz}$\cite{DOHERTY20131} originates from spin-spin interaction and the Hamiltonian of the spin triplet is
\begin{equation}\label{eq:NVHamiltonian1}
\hat{H}_0=D\hat{S}_z^2+g\mu_B\bm{B}_0\cdot\bm{S},
\end{equation}
where the second term accounts for the Zeeman splitting with an external magnetic field $\bm{B}_0=(B_x,B_y,B_z)$. The spin vector $\bm{S}=(\hat{S}_x,\hat{S}_y,\hat{S}_z)$ contains the ($S=1$) spin operators $\hat{S}_i$, which are defined in the basis of the $m_S=0,-1,1$ spin projections corresponding to the $\hat{S}_z$ eigenstates 
$\lbrace|0\rangle,|-\rangle$ and $|+\rangle\rbrace$.
In cases where the magnetic field is not aligned with the NV-axis $\hat{c}$, the Hamiltonian \eqref{eq:NVHamiltonian1} can be written as
\begin{equation}\label{eq:NVHamiltonian2}
\hat{H}_0=D\hat{S}_z^2+g\mu_B B_0(\sin \theta_B~\hat{S}_x+\cos \theta_B~\hat{S}_z),
\end{equation}
where the relative angle $\theta_B$ quantifies the orientation of the external magnetic field in the frame of the NV center. Since the Hamiltonian \eqref{eq:NVHamiltonian2} depends explicitly on $\theta_B$, the according eigenenergies and, consequently, the frequencies $\omega_{\theta_B}^\pm$ of the dipole transitions $|0\rangle\leftrightarrow|\pm\rangle$ vary for different orientations of the magnetic field with respect to the NV-axis. Hence, the transition spectrum of a single NV center consists of two frequency branches $\omega_{\theta_B}^\pm$, whose behavior with respect to varying bias field $B_0$ is determined by the given value of $\theta_B$. 
\subsection{NV ensemble in single crystal diamond}
In case of an NV ensemble every single of the numerous contained NV centers contributes two branches to the transition spectrum. If the single NV centers were randomly oriented, all branches would combine to a blurred spectrum. But due to the particular symmetry of the diamond crystal, the actual transition spectrum of an NV ensemble inside a single crystal consists of single branches. 
For a single crystal diamond, the lattice has a fixed orientation in the lab frame, but inside the diamond structure the actual orientation of an NV center is not controllable during fabrication. Thus, the NV axis can be aligned along the four crystal directions shown in Fig. \ref{fig:NVCrystalAngle}$(a)$, which are at a tetrahedral angle of $\theta_t=\SI{109.5}{\degree}$ to each other. A single nanodiamond contains NV centers with orientations equally distributed over these four directions. Accordingly, for a fixed $\bm{B}_0$-field orientation, the magnetic field is aligned at a different angle with respect to each of these four directions. Consequently, there are not numerous overlaying resonance branches, but each of the four possible NV-axis orientations gives rise to two resonances and in total there are eight resonances for an NV ensemble.

In order to describe the expected magnetic resonances with a formula, the NV center orientation is expressed in terms of the relative angles $\theta_{Bi}$ between the possible NV-axis and the magnetic field as shown in the Fig. \ref{fig:NVCrystalAngle}$(b)$. Choosing one of the NV-axes, $\hat{c}_0$, to define the $z$-direction and one of the carbon atoms to lie on the $x$-axis, the normalized magnetic field vector and the four NV-axes $\hat{c}_i$ are parameterized in spherical coordinates as $\hat{\bm{B}}=(
\sin\theta_B\cos\varphi_B,\sin\theta_B\cos\varphi_B,\cos\theta_B)$ and $
\hat{c}_i=(
\sin\theta_i\cos\varphi_i,\sin\theta_i\cos\varphi_i,\cos\theta_i)$ with $\theta_0=0$, $\theta_{1,2,3}=\theta_t$, $\varphi_{0,1}=0$ and $\varphi_{2,3}=2\pi/3,4\pi/3$.
The introduced coordinate system is thus fixed to the crystal lattice. Based on these vectors, the angles $\theta_{Bi}$ between the single NV-axes and the magnetic field in Fig. \ref{fig:NVCrystalAngle} are related to the corresponding scalar product as
\begin{equation}
\cos\theta_{Bi} =\hat{\bm{B}}\cdot\hat{c}_i=\sin\theta_B\sin\theta_i\cos(\varphi_B-\varphi_i) +\cos\theta_B\cos\theta_i.
\end{equation}
Inserting the angle $\theta_{Bi}$ obtained from this expression into the Hamiltonian for non-aligned magnetic field \eqref{eq:NVHamiltonian2} and calculating the transition frequencies between the corresponding eigenenergies yields the spectra depending on the polar as well as the azimuth angle of the magnetic field as plotted  with respect to the strength of the applied magnetic field in Fig. \ref{fig:NVCrystalAngle}$(c)$.
A comparison of the two plots corresponding to different parameter sets $(\theta_B,\varphi_B)$ highlights the sensitivity of the resonance spectrum of single crystal diamond hosting NV centers at various orientations to rotations of the crystal in a magnetic field. In that way, information about the spatial orientations can be extracted by exploiting the existing relation.

From experimental data of the eight different magnetic resonance frequencies in a nanodiamond of known crystal orientation, the angles $\theta_B$ and $\varphi_B$ can be calculated or, in turn, from a known magnetic field alignment, the crystal directions of a nanodiamond can be derived. Hence, the resonance experiment can be used to sense a probe magnetic field resolving its direction on the one hand and to measure the crystal structure of a nanodiamond indirectly on the other hand.\cite{PhysRevA.96.042115, 1367-2630-16-6-063067}
\subsection{Spin dynamics}
In order to model the quantum mechanical coupling of an NV ensemble to the calculated spin wave field \eqref{eq:C1(x)Result}, the corresponding Bloch equations are solved. Therefore, we use a five level scheme similar to \cite{FiveLevel}, which includes the NV energy levels relevant for the processes in an ODMR experiment. The NV is pumped by a continuous laser, which excites the system from its ground state $^3$A$_2$ triplet $\lbrace|0\rangle,|\pm 1\rangle\rbrace$ to the excited state $^3$E triplet $\lbrace|e0\rangle,|\pm e1\rangle\rbrace$. From the excited triplet states, the system partially decays optically back to the ground state manifold and the intensity of the emitted fluorescent photons is detected in ODMR experiments.

Without external microwave drive and in the presence of optical excitation the fluorescence would just have a constant intensity and the system would be polarized in the $m_s=0$. 
due to the inter-system crossing channel via the excited singlet $^1$A$_1$ state $|s\rangle$. The induced resonance microwave transitions within the ground state manifold pump the system out of $m_s=0$ state and in the presence of optical excitation lead to a change of fluorescence intensity. The intensity in fact drops because now the system spends more time in the $|s\rangle$ state, which is not optically active at the same photon frequency.

The dynamics of this mechanism are described by the time-dependent five-dimensional density matrix $\rho(t)$ of the system written in the basis of $\lbrace|0\rangle,|1\rangle,|e0\rangle,|e1\rangle,|s\rangle\rbrace$. Thereby, only the driven $m_S=1$ triplet sublevels are regarded, since we neglect transitions from the singlet to the spin state that is not driven by the microwave field. When the transition into the $m_S=-1$ state is driven, we use the replacement $|1\rangle\rightarrow|-1\rangle$. The density matrix is determined by solving the Master equation in Lindblad form,
\begin{equation}\label{eq:LindbladEquation}
\partial_t\rho=i[\rho,\hat{H}]+\sum_{\mu>0}\left[L_\mu\rho L_\mu^\dagger-\frac{1}{2}L_\mu^\dagger L_\mu\rho-\frac{1}{2}\rho L_\mu^\dagger L_\mu\right], 
\end{equation}
where the operators $L_\mu$ describe a non-unitary time evolution due to dissipative interactions between the system and the environment, which is given by the photon bath of the pumping laser and the emitted fluorescence. 
For the special case of the five level system, the Hamiltonian $\hat{H}$ for the closed system in \eqref{eq:LindbladEquation} is decomposed into the static part and the interaction part, $\hat{H}=\hat{H}_0+\hat{H}_I$. The static Hamiltonian $\hat{H}_0$ is diagonal containing the eigenenergies $\varepsilon_i$ of the five states $|i\rangle$ with $i=0,1,e0,e1,s$. Writing the classical microwave field as $\bm{B}_\text{mw}(t)=\bm{B}_\text{mw}\cos(\omega t)$, the interaction Hamiltonian is given by 
\begin{equation}
\label{eq:HamiltonianClosed}
\hat{H}_I=\frac{\Omega_R}{2} \left(e^{i\omega t}+e^{-i\omega t}\right)\left(|1\rangle\langle 0|+|0\rangle\langle 1|\right),
\end{equation}
with the Rabi frequency $\Omega_R=\gamma\braket{0|\bm{B}_\text{mw}\cdot\bm{S}|1}$ denoting the coupling strength due to magnetic dipole interaction between the driving field and the ground state triplet. Here, the spin vector $\bm{S}=1/2\bm{\sigma}$ can be represented by the Pauli matrices $\bm{\sigma}=(\sigma_x,\sigma_y,\sigma_z)$. The operators $L_\mu$ in \eqref{eq:LindbladEquation} represent the various dissipation processes as well as the continuous pumping, which excites the system from the ground state manifold to the excited state manifold at pumping rate $\Gamma_p$, where the $m_S$ quantum number is left unchanged, i. e. $L_p^0=\Gamma_p^{1/2}|e0\rangle\langle 0|$ and $L_p^1=\Gamma_p^{1/2}|e1\rangle\langle 1|$. The inverse process, the direct decay back into the ground states, happens at rate $\Gamma_0$ and the corresponding operators are $L_0^0=\Gamma_0^{1/2}|0\rangle\langle e0|$ and $L_0^1=\Gamma_0^{1/2}|1\rangle\langle e1|$. For the inter-system crossing the coupling between the lower level $|e0\rangle$ and $|s\rangle$ is much smaller than the coupling between the highest level $|e1\rangle$ and $|s\rangle$, therefore, the latter is neglected. Hence, the corresponding operator is $L_{es}=\Gamma_{es}^{1/2}|s\rangle\langle e1|$. The second inter-system crossing from $|s\rangle$ to the ground state triplet $|0\rangle$ and $|1\rangle$ is modeled to have the same probabilities for ending up in the final states $|0\rangle$ and $|1\rangle$ with the operators $L_{sg}^0=(\Gamma_{sg}/2)^{1/2}|0\rangle\langle s|$ and $L_{sg}^1=(\Gamma_{sg}/2)^{1/2}|1\rangle\langle s|$. Further, the $T_1$-decay at rate $\Gamma_1=1/T_1$ from the $m_S=\pm1$ and $m_S=0$ ground states into the equilibrium, which is assumed to lie at equally populated states is included. We also treat the transverse relaxation including dephasing inside the ground state spin triplet at rate $\Gamma_2$. The corresponding operators can be written in terms of spin operators as $L_1=(\Gamma_1/2)^{1/2}\left(|0\rangle\langle 1|+|1\rangle\langle 0|\right)$ and $L_2=(\Gamma_2/2)^{1/2}~\hat{\sigma}_z$. Inserting these operators and the Hamiltonians $\hat{H}_0$ and $\hat{H}_I$ into the Master equation \eqref{eq:LindbladEquation} yields the time derivative of the density matrix $\rho$ in the form of 15 coupled first order differential equations with time-dependent coefficients. Since some of these equations are redundant, a system of only seven differential equations actually has to be solved, where the rotating wave approximation is used. In the special case of continuous pumping a single measurement of the ODMR contrast at fixed driving frequency $\omega$ is detected over a time long enough to allow the system to evolve into a dynamical equilibrium. Hence, the solution of interest is a steady where the time derivative in \eqref{eq:LindbladEquation} can be set to zero. Finally, the matrix elements $\rho_{ij}$ are obtained by solving the remaining system of linear equations.
\subsection{Optical detection}
\begin{figure*}
\includegraphics[width=0.5\textwidth]{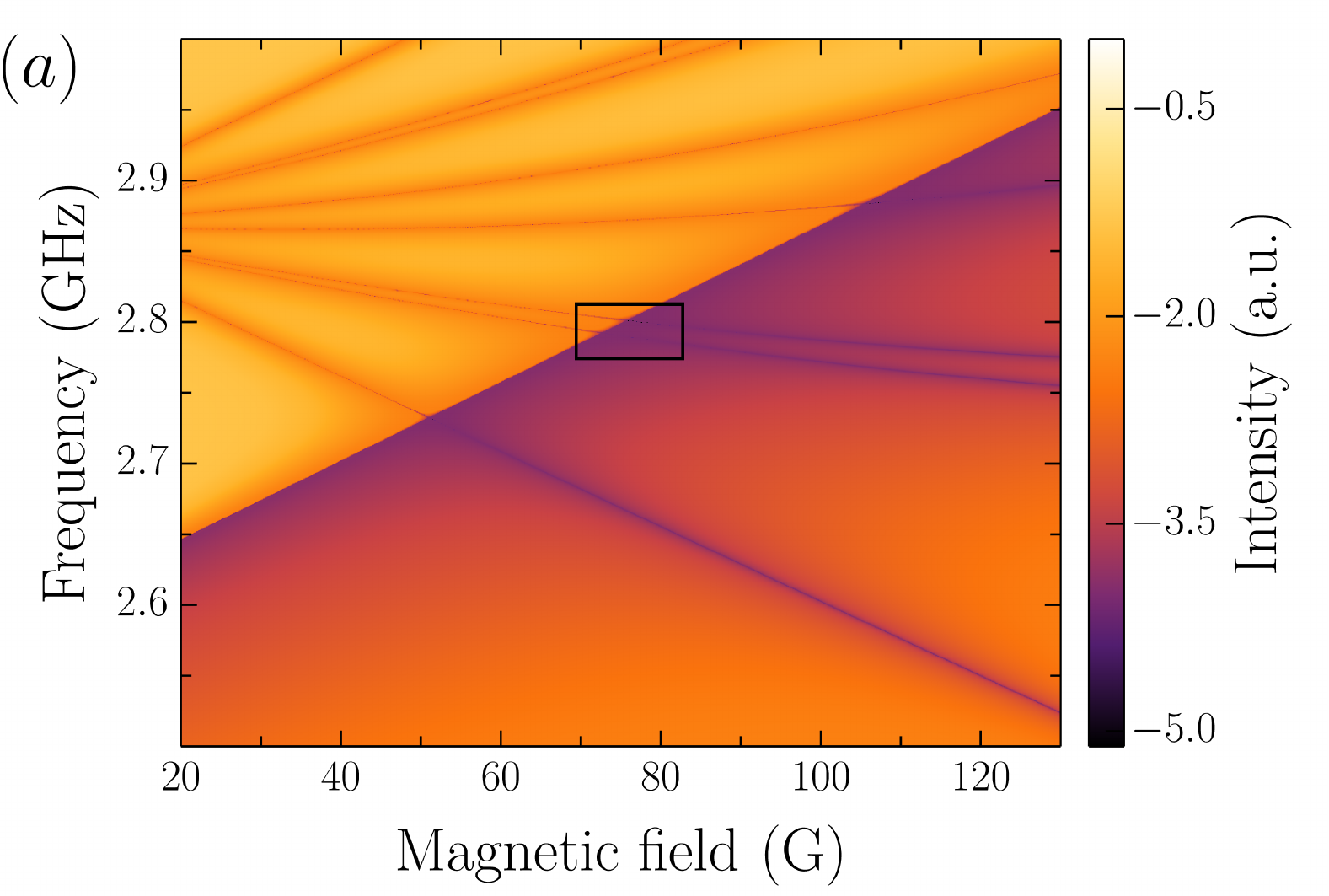}\includegraphics[width=0.5\textwidth]{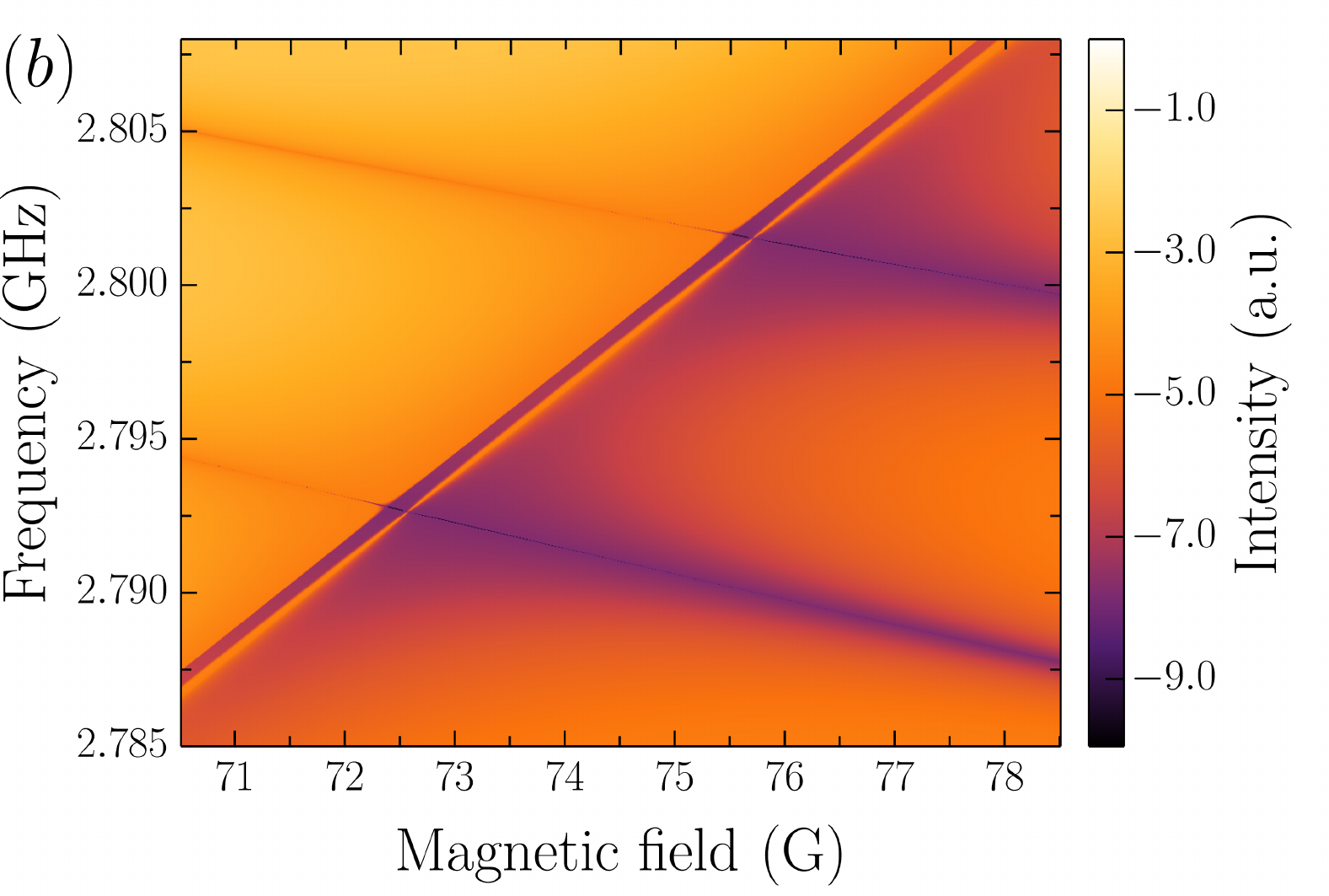}
   \caption{$(a)$ and $(b)$: ODMR intensity $I$ depending on the magnetic field and the driving frequency. The used parameters correspond to a nanodiamond at $x_\text{NV}=\SI{5}{\micro\meter}$ away from the antenna and a driving power of $\SI{0.5}{\micro\watt}$. For the magnetic field orientation the arbitrary values $\theta_B=\SI{14}{\degree}$ and $\varphi_B=\SI{7}{\degree}$ are chosen. In $(b)$, a zoom of the black boxed area in $(a)$ is shown, which highlights the substructure of the resonances right at the cut-off frequency $\omega_\text{sw}^\text{max}$.}
\label{fig:SpectrumZoom}
\end{figure*}
The ODMR intensity $I$ can be calculated from the obtained solution for the density matrix $\rho$. It is given by the number of emitted photons, when the excited states $|e0\rangle$ and $|e1\rangle$ decay optically to the ground state triplet. Due to the existence of the inter-system crossing channel, only a part of the population in $|e1\rangle$ contributes to the intensity. In detail, only the fraction $\Gamma_0/(\Gamma_0+\Gamma_{es})$ remains within the spin triplet channel. Treating the detected intensity to be proportional to these effective probabilities, it can be written as
\begin{equation}
I\propto\rho_{e0e0}+\frac{\Gamma_0}{\Gamma_0+\Gamma_{es}}\rho_{e1e1}.
\end{equation}
For the steady state solutions, the normalized intensity depending on the detuning $\delta=\varepsilon_1-\omega$ turns out to be Lorentzian,
\begin{equation}\label{eq:Lorentzian}
I(\delta)=\frac{\delta^2}{\delta^2+\widetilde{\gamma}^2},
\end{equation}
where $\widetilde{\gamma}$ describes the width of a resonance centered at $\delta=0$, which is simplified assuming the following. First, the decay rate from the excited state $|e1\rangle$ into the triplet ground state and the inter-system crossing channel are of the same order, i. e. $\Gamma_{es}\approx\Gamma_{0}$. Second, the ground state triplet is not completely depopulated during the experiment, which requires a pumping rate much lower than the decay rate, $\Gamma_p\ll\Gamma_0$, and the $T_1$-decay being much slower than the
leakage via the inter-system crossing channel, $\Gamma_1\ll\Gamma_{sg}$.\cite{FiveLevel} Under these conditions, the parameter $\widetilde{\gamma}$ takes the compact form
\begin{equation}\label{eq:Width}
\widetilde{\gamma}=\sqrt{\left(2\Gamma_2^\text{eff}\right)^2+ \frac{4\Gamma_2^\text{eff}(1+\Gamma_p/(4\Gamma_{sg}))}{\Gamma_1+\Gamma_p/4}\Omega_R^2},
\end{equation}
with the effective transverse relaxation rate $\Gamma^\text{eff}_2=\Gamma_2+\Gamma_p+\Gamma_1/2$. So far, we assumed degenerate $m_S=\pm1$ sublevels. But since the NV ensemble in the considered setup is placed in a static bias magnetic field, these sublevels are split by the corresponding Zeeman energy. Thus, if the NV ensemble is driven with a microwave field at frequency $\omega$, each of the transitions $|0\rangle\leftrightarrow|+\rangle$ and $|0\rangle\leftrightarrow|-\rangle$ will be excited with different detunings $\delta_\pm=\omega_\pm-\omega$. In addition, the resonance spectrum of a single nanodiamond is a combination of the resonances originating from four different alignments of the NV-axes as discussed for an NV ensemble. To calculate the ODMR spectrum of an NV ensemble inside a nanodiamond with eight occurring resonances, we assume the eight expected resonances being far off-resonant with each other. Hence, at $\delta_+^i=0$ the detuning from the second resonance $\delta_-^i$ is large enough and single transitions are excited separately. Therefore, the contributions $I(\delta_\pm^i)$ in \eqref{eq:Lorentzian} from each of the eight resonances are summed up to the intensity of ODMR fluorescence,
\begin{equation}\label{eq:IntensitySum}
I(\omega)=\sum_{j=\pm}\sum_{i=1}^{4}\frac{(\omega_j^i-\omega)^2}{(\omega_j^i-\omega)^2+\gamma^2}.
\end{equation}\setcitestyle{super}
Due to the large magnetic field amplitudes of the driving field in the experiment, the line width $\widetilde{\gamma}$ in \eqref{eq:Width} is dominated by the second term, which means that the lines can be assumed to be purely power broadened\cite{FiveLevel}. 
In the case of low pumping power, the condition $\Gamma_p/4\ll\Gamma_{sg}$ is fulfilled and $\widetilde{\gamma}$ can be further approximated by
\begin{equation}\label{eq:WidthApprox}
\widetilde{\gamma}\approx\sqrt{ \frac{4~\Gamma_2^\text{eff}}{\Gamma_1+\Gamma_p/4}}~\Omega_R.
\end{equation}
For a plot of the ODMR intensity $I(\omega)$ in dependence of the bias field strength, we assume realistic values for the relaxation times $T_1=1/\Gamma_1$ and $T_2=1/\Gamma_2$ of the NV center to lie in the range of ms\cite{T1} and $\si{\micro\second}$\cite{T2} and a pumping rate $\Gamma_p\approx 10^6\si{\per\second}$, so that the line width is approximately $2\Omega_R$.

Based on the intensity \eqref{eq:IntensitySum} with the simplified line width \eqref{eq:WidthApprox} and using the spin wave field amplitude \eqref{eq:C1(x)Result} calculated in section \ref{Sec3}, a theoretical ODMR spectrum is plotted in Fig. \ref{fig:SpectrumZoom}. In the frequency regime above the cut-off of spin wave excitation, i. e. $\omega>\omega_\text{sw}^\text{max}$, we use a microwave driving field amplitude corresponding to a pure antenna field $\bm{H}_\text{ant}$ without spin wave excitation inside the YIG film.

The parameters $x_\text{NV}$ and $P$ entering the function for $\bm{H}_\text{sw/ant}$ are chosen to be equal to the experimental spectrum in Ref.~\onlinecite{Andrich} (Supplementary), which are given by the distance between the nanodiamond and the antenna $x_\text{NV}\approx \SI{5}{\micro\meter}$ and a driving power of $P=\SI{0.5}{\micro\watt}$. The angle of the bias field is chosen arbitrary to be $\theta_B=\SI{14}{\degree}$ and $\varphi_B=\SI{7}{\degree}$. The zoom of two resonances within the black boxed area in $(a)$ right at the maximal spin wave excitation frequency is given in Fig. \ref{fig:SpectrumZoom}$(b)$. The more the driving frequency abroaches the cut-off frequency $\omega_\text{sw}^\text{max}$, the stronger the resonances are broadened, which results from the maximum of the amplitude function $H^\text{I}$ at the upper frequency limit. In addition, the high resolution in Fig. \ref{fig:SpectrumZoom}$(b)$ also shows a sharp peak in the intensity apparent as a bright line parallel to the border between the driving regimes. The reason for this intensity increase is the substructure of the amplitude function in the spin wave driving regime and it corresponds to the first dip in Fig. \ref{fig:C1Plot} at frequency $\widetilde{\omega}_\text{sw}$. Between this peak and the cut-off frequency, the spin wave field takes a maximal value right before it drops to zero. This becomes apparent by the dark line in Fig. \ref{fig:SpectrumZoom}$(b)$ crossing the resonance branches from the bottem left to the top right and a similar dark line occurs in experimental ODMR spectra\cite{Andrich}.
This consistency between the experiment and the theory plot suggests, that the amplitude function \eqref{eq:C1(x)Result} describes the field amplitude in the spin wave regime at a satisfying level of accuracy.
\begin{figure*}
\includegraphics[height=0.25\textheight]{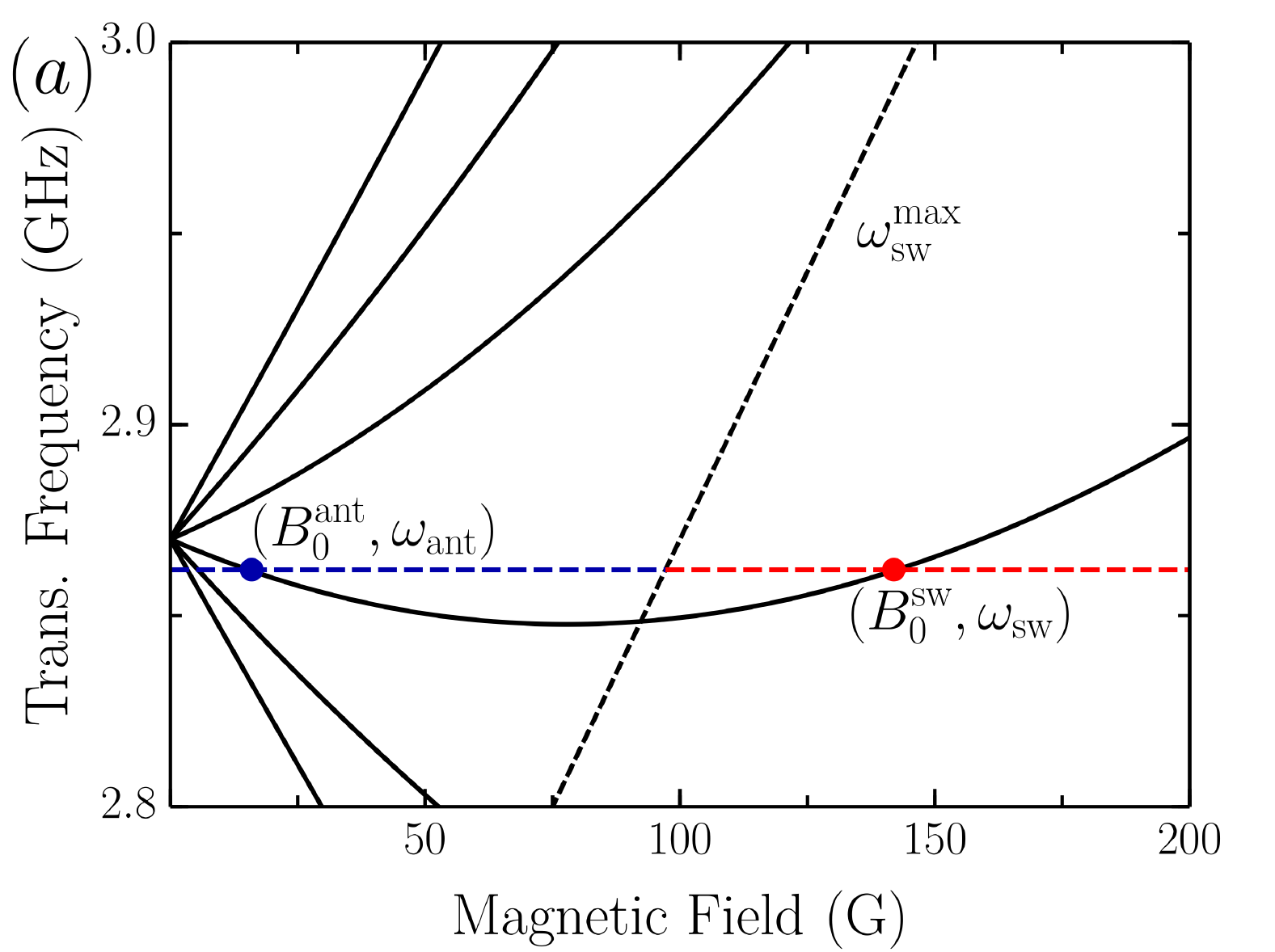}
\includegraphics[height=0.25\textheight]{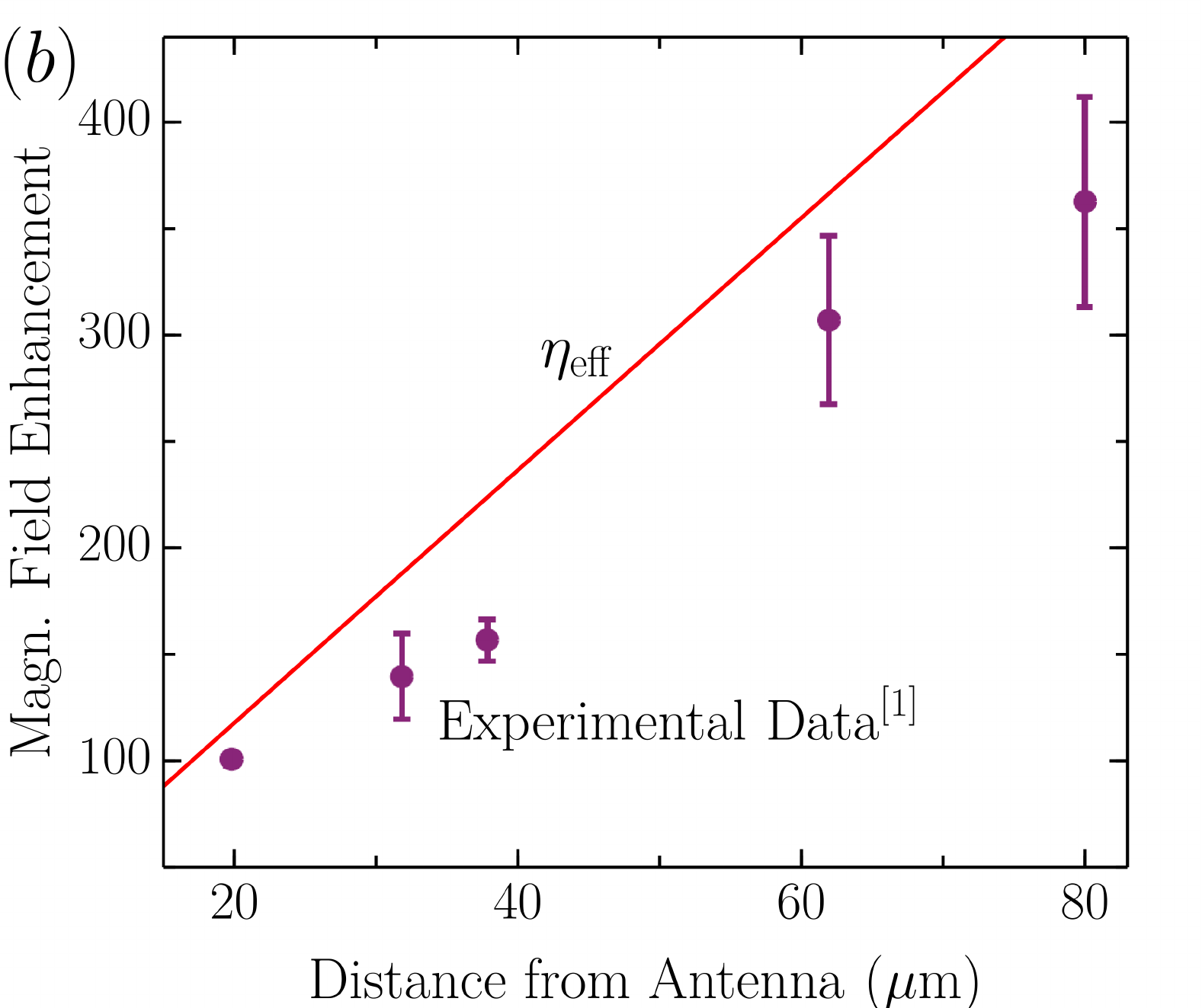}
\caption{$(a)$ Plot of the resonance frequencies of an NV center (black) depending on the bias field. The  NV-axis is oriented in the $xy$-plane at an angle of $\SI{78}{\degree}$. The black dashed line represents the step between spin wave and antenna regime. The intersections of the driving frequency (blue/red dashed) with one of the branches indicates possible measurement settings. $(b)$ Dependence of the effective enhancement $\eta_\text{eff}$ on the $x$-component of the coupling NV ensemble. The theoretical model corresponds to the red line and the data points represent the experimental results of Andrich \textit{et al.} presented in Ref.~\onlinecite{Andrich}.\label{fig:Enhancement}}
\end{figure*}

\section{\label{Sec5}Enhancement}
Due to the presence of spin waves inside the magnetic film, the magnetic field coupling to an NV ensemble is enhanced compared to the amplitude of a pure antenna field. Moreover, the spin wave field has the great advantage of higher decay length, which becomes apparent when inspecting the explicit dependencies of the spin wave field in \eqref{eq:C1(x)Result}.
The lack of $x$-dependence of the field amplitude indicates a very important difference from a pure antenna driving field. The microwave field of the antenna decays as $1/x$, so that the coupling strongly depends on the position of the interacting NV center. In a system of defect spins in a quantum system addressed or read out with an oscillating magnetic field, information can be transferred over a distance, which increases with the length over which the field amplitude is conserved. Hence, the $x$-independent spin wave field amplitude in \eqref{eq:C1(x)Result} appears to lead to a constant coupling over a large distance. Realistically, the coupling decays due to the neglected exchange energy terms in the susceptibility tensor. Regarding the corresponding terms would give rise to a magnetic amplitude damping in \eqref{eq:C1(x)Result} along the $x$-coordinate. But YIG has a particularly low damping parameter and the excited spin waves propagate at low $k$, thus justifying the made approximation. Along the $z$-direction, $\tilde{C}_1$ decays exponentially with the distance to the surface, i.e. the spin wave modes are confined to the surface of the thin film. This guarantees a highly localized field, so that the effect of spin wave enhancement is only present in the vicinity to the magnetic film. 

Using the obtained result for the spin wave field allows us to investigate the enhancement of the coupling strength $\Omega_R$ with respect to the case without spin wave excitation. The enhancement factor is defined as $\eta=\Omega^\text{sw}_R/\Omega_R^\text{ant}$, where $\Omega^\text{sw}_R$ denotes the Rabi frequency in the case of spin wave excitation and $\Omega^\text{ant}_R$ is the same quantity for the pure antenna driving without any back action originating from magnetic material. More specifically we calculate the antenna field without regarding the stray field of the YIG, which is reasonable in the regime without spin wave excitation, since the thin film is located in an approximately homogeneous magnetic field and due to its small height the influence on the antenna field becomes negligible. Both quantities, the spin wave and the antenna field, depend on the angle between the respective field and the NV center, which is non-zero in the general case.

In a coordinate system $S'$ with $z'$-axis parallel to the NV-axis $\hat{c}$ and the antenna field lying in the $x'z'$-plane, the magnetic fields of antenna and spin waves are parametrized as $\bm{B}_\text{ant}=B_\text{ant}(
\sin\theta_\text{ant}, 0,\cos\theta_\text{ant})$ and $\bm{B}_\text{sw}=B_\text{sw}(\sin\theta_\text{sw}~\cos\varphi_\text{sw}, \sin\theta_\text{sw}~\sin\varphi_\text{sw},\cos\theta_\text{sw})$, where $\theta_\text{ant/sw}$ denotes the angle with respect to the $\hat{c}$-axis and the spin wave field is at the angle $\varphi_\text{sw}$ with respect to the $x'z'$-plane. The Rabi frequency scales with the off-diagonal terms in $\bm{B}_\text{sw/ant}\cdot\bm{S}$, so that $\Omega_R^\text{sw}\propto B_\text{sw}\sin \theta_\text{sw} ~\exp(-i\varphi_\text{sw})$ and analogously for the antenna driving field. Inserting this into the enhancement ratio yields
\begin{equation}\label{eq:AngleS'}
\eta=\frac{H_\text{sw}}{H_\text{ant}}~\frac{\sin \theta_\text{sw}}{\sin \theta_\text{ant}}~e^{-i\varphi_\text{sw}}.
\end{equation}
Since the phase factor $e^{-i\varphi_\text{sw}}$ does not affect the absolute value of this enhancement, it is omitted in the following. Furthermore, we want to adapt our result to the experimental method described in \cite{Andrich}, where the resonance between the same levels is measured twice in the different driving regimes, the antenna driving and the spin wave driving regime. To measure both regimes at the same frequency, a resonance has to lie once above the maximal excitation frequency of spin waves and once below, so that the different regimes can be tuned by the bias magnetic field.
For the realization of such a measurement, the orientation of the bias field has to be non-collinear to the NV-axis, whereby the states $|0\rangle$ and $|\pm \rangle$ become coupled and the energy splitting does not depend linearly on the magnetic field. Thus, the resonance branches are tilted as plotted in Fig. \ref{fig:Enhancement}$(a)$. The plot shows the NV resonances in the case of the NV-axis being aligned in the $xy$-plane at an angle of $\theta_B=\SI{78}{\degree}$ with respect to the bias magnetic field. In Fig. \ref{fig:Enhancement}$(a)$ only six resonances are visible, because the magnetic field encloses the same angle with two of the possible crystal directions leading to these resonances being two-fold degenerate. An excitation frequency of $\SI{2.86}{\giga\hertz}$ is on resonance with a bias field $B_{0\text{ant}}$ in the antenna driving regime, which is marked by the blue dot in Fig. \ref{fig:Enhancement}$(a)$. A second resonance (red dot) occurs in the regime of spin wave excitation at $B_{0\text{sw}}$.
For an optimal resolution of the detection in both regimes, the antenna current, i. e. the driving power $P$ is varied. The antenna field decays rapidly with the distance of the NV center from the antenna, so that a driving power of the order of $\si{\milli\watt}$ is required to obtain a clear signal. If the same driving power was used in the spin wave driving regime, the resonance would be significantly power broadened and hence, a lower power at several $\si{\micro\watt}$ is chosen. Hence, the enhancement ratio $\eta$ has to be corrected by the impact of varying driving power.
Both, the antenna field amplitude $H_\text{ant}$ and the spin wave field amplitude $H_\text{sw}$ depend linearly on the driving current $I_0=(2P/Z)^{1/2}$, which is an effective ac current through a wire with impedance $Z$. Finally, the effective enhancement ratio is obtained by normalizing the fields with respect to the driving power $P_\text{ant/sw}$,
\begin{widetext}
\begin{equation}\label{eq:EnhancementAngle}
\eta_\text{eff}=\sqrt{\frac{P_\text{ant}}{P_\text{sw}}}~\frac{H_\text{sw}}{H_\text{ant}}~\frac{\sin\theta_\text{sw}}{\sin\theta_\text{ant}}
=4\sqrt{2}\pi~\sqrt{\frac{P_\text{ant}}{P_\text{sw}}}~\frac{\frac{\sin(k'_xw/2)}{ k'_x}\left(\frac{\omega_M(\omega_0+\omega_M+\omega)}{4\omega^2-(2\omega_0+\omega_M)^2}~e^{-2k'_xd}+\frac{\omega_0+\omega_M-\omega}{2\omega_0+\omega_M-2\omega}\right)e^{-k'_xz}}{\ln\left(\left(\frac{w}{2}-x\right)^2/\left(\frac{w}{2}+x\right)^2\right)}~\frac{\sin\theta_\text{sw}}{\sin\theta_\text{ant}}.
\end{equation}
\end{widetext}

With this expression, we are able to make a theoretical prediction for the enhancement for an NV center located at $x_\text{NV}$. Since the far field approximation was made, the condition $x\gg k^{-1}$ has to be fulfilled and for wave numbers of the excited spin waves\cite{Andrich} of the order of $\sim 10^5\si{\per\meter}$, the expression \eqref{eq:EnhancementAngle} applies for distances above $\sim \SI{10}{\micro\meter}$. For example, the resonances of a nanodiamond at $x_\text{NV}=\SI{20}{\micro\meter}$ driven by a field at $\SI{2.862}{\giga\hertz}$ occur in the antenna regime at $\SI{15}{\gauss}$ and in the spin wave driving regime at $\SI{145}{\gauss}$.
Using these parameters, the enhancement factor of the coupling between the defect spin and the driving field is
$\eta_\text{eff}\approx 117$, 
which is close to the experimentally measured enhancement  $\eta_\text{eff}\approx 100$ detected via Rabi experiments\cite{Andrich}. Moreover, when distance $x$ in expression \eqref{eq:EnhancementAngle} is varied, the enhancement factor $\eta_\text{eff}$ behaves as plotted in Fig. \ref{fig:Enhancement}$(b)$, along with the experimental data of Andrich \textit{et al.} \cite{Andrich}. The theoretical prediction matches the experimental behavior, although the experimental data are slightly shifted to lower enhancement values with respect to the predicted theory curve.
Nevertheless, the linear $x$-dependence of  $\eta_\text{eff}$ 
not only shows good agreement with the trend of experimental data, but also an achievable enhancement factor of more than 400 at distances above  $70\,\mu{\rm m}$ 
 which underscores the great advantage of spin wave excitation inside the magnetic film. For low magnetic damping the spin waves do not decay as fast as the microwave field leading to a large enhancement far away from the antenna.

\section{\label{Sec6}Conclusion}
We have presented a theoretical model that provides a complete description of real experimental hybrid systems consisting of driven spin waves coupled to NV spins. All results are obtained analytically and especially the description of spin wave modes follows a fundamental approach from the basics of Maxwell's equations. In that way, we derive an analytical solution for the magnetic field amplitude, which allows for a theoretical calculation of the coupling to an NV ensemble. Therefore, we regard every possible spatial orientation of an NV center in a real nano-diamond, thus obtaining the full spectrum capturing all eight resonances.
Our theoretical intensity plots show good agreement with existing experimental data\cite{Andrich}, what clearly highlights the suitability of the developed model. The existence of the substructure and an upper frequency bound limiting the spin wave driving regime are important features, which follow from the theoretical calculations. They provide an accurate but at the same time very applicable method of tuning the drive of spin rotations of an NV center, where the resonant coupling below the cut-off frequency in the spin wave driving regime are calculated to be much stronger with respect to the case of pure antenna driving. This difference in the coupling strength is a powerful tool to speed up qubit operations in the field of quantum computing. The derived enhancement factor $\eta_\text{eff}$ is strongly dependent on the distance between the source of the microwave field and the NV ensemble. Accordingly, the magnetic field at the NV spin can be enhanced by two orders of magnitude and the predicted behavior for longer distances is in good agreement with recent experiments\cite{Andrich} indicating the strength of our model as well as the applicability of the related assumptions and approximations. The presented solutions are obtained using the far field approximation, which holds for distances above $\sim \SI{10}{\micro\meter}$. This regime is of particular interest for technical realizations, since it is a general problem to couple a sensitive quantum system to a microwave antenna across distances of more than a few $\si{\micro\meter}$ in view of the rapidly decaying antenna field. In contrast, we show that the spin wave field is very stable across the distance enabling a coupling to NV centers far away from the antenna by using a magnetic thin film.

If the enhancing effect over a long distance is of particular interest the made approximations are reasonable and the model provides good results. But in case of a modified system operating at a different conditions, possibly the limits of our theoretical model could be reached. If a nanodiamond is positioned very close to the antenna, there is need of a solution outside the far field regime, which is applicable for short distances. An exemplary real system would be an array of nanodiamonds, where the antenna is not grown besides but in between the array, so that the NV centers are partly in the near field and partly in the far field regime. Strictly speaking, this would require a mid field solution as well.

Further unconsidered effects, which had to be taken into account in order to optimize the presented model, are additional boundary effects caused by the sourrounding material being non-perfect magnetic vaccum or spin wave reflections in case of the film having finite expansion in the $xy$-plane. A refinement of the model considering those aspects would lead to important insights with respect to new applications of a spin wave mediated coupling. Since the spectrum of the spin waves is mainly determined by the shape of the ferromagnetic sample, the driven spin wave modes could be adapted by shaping, for example, a ferromagnetic sphere or a disk. In this way, the technique of the experiment of Andrich \textit{et al.}\cite{Andrich} could be transferred to various magnetic resonance experiments by changing the driven spin wave modes to lie on resonance with other defect spins like different color centers in diamond or vacancies in silicon carbide. The usage of a spin wave resonator, which has already been experimentally realized\cite{doi:10.1063/1.4766918}, would lead to an additional improvement of spin wave excitation. This provides an outlook towards the possible long-term goal of achieving a coherent coupling between single magnons and single defect spins, thus entering the quantum regime.

\bibliography{literatureCompile}

\end{document}